\documentclass[twocolumn,prl]{revtex4}
\usepackage{graphicx}
\usepackage{amssymb,amsmath,bm}
\newcommand{\be}{\begin{equation}}
\newcommand{\ee}{\end{equation}}
\newcommand{\bea}{\begin{eqnarray}}
\newcommand{\eea}{\end{eqnarray}}

\newcommand{\p}{\partial}

\newcommand{\lp}{\left(}
\newcommand{\rp}{\right)}

\renewcommand{\phi}{\varphi}
\renewcommand{\epsilon}{\varepsilon}
\renewcommand{\vec}[1]{{\bf #1}}

\renewcommand{\Im}{\mathop{\rm Im}\nolimits}

\begin{document}

\title{Transport in Graphene p-n Junctions in Magnetic Field}
\author{
A. V. Shytov,${}^1$ Nan Gu,${}^2$ L. S. Levitov${}^2$}
\affiliation{
${}^1$
Brookhaven National Laboratory, Upton,
New York 11973-5000\\
${}^2$ Department of Physics,
 Massachusetts Institute of Technology, 77 Massachusetts Ave,
 Cambridge, MA 02139
}

\begin{abstract}
Ballistic transport in graphene p-n junctions in the presence of magnetic
field exhibits two distinct regimes: At low fields, transport is partially
suppressed by the field. When the field exceeds a certain critical value, the
junction is pinched off by the Landau level formation.
Transmission
and conductance are found in the entire range of fields using Lorentz boost
and mapping to the Landau-Zener problem.
We show that perfect transmission occurs at a field-dependent collimation
angle, indicating that the chiral dynamics of massless Dirac fermions
persists at a finite magnetic field.
A current switch, utilizing field-tunable collimation angle,
is proposed. With  
a generalization of the developed approach we study 
transmission through p-n junctions in graphene bilayer.
\end{abstract}
\maketitle

Graphene p-n junctions, fabricated recently
in locally gated samples\,\cite{Huard07,Williams07,Ozyilmaz07}, 
provide a new tool to study electron transport.
Charge carriers in graphene
mimic relativistic Dirac 
particles with zero mass and linear dispersion
relation $\epsilon=\pm v_F|\vec p|$ with $v_F\approx 10^8\,{\rm cm/s}$. 
Graphene p-n junctions are predicted to exhibit signatures of chiral dynamics of massless Dirac particles:
perfect transmission normal 
to the junction\,\cite{Katsnelson06b,KatsnelsonSSC} 
and collimation
of the transmitted particles\,\cite{Cheianov06}.
Ballistic transport in 
p-n junctions was proposed as a means to realize 
an electron lens\,\cite{Cheianov07}.

The properties of the p-n-p system studied in Ref.\,\cite{Huard07} 
could be mainly explained by conduction in the disordered p and n regions, 
rather than in the p-n junctions.
Likewise, the effects 
in quantizing magnetic fields \cite{Williams07,Ozyilmaz07} 
were understood
from edge state transport in the p and n regions,
with the p-n 
interface merely providing mode mixing\,\cite{Abanin07}. 
In neither of the experiments \cite{Huard07,Williams07,Ozyilmaz07} 
the effects of ballistic
transmission\,\cite{Katsnelson06b,Cheianov06,Cheianov07} 
seemed to stand out. 

This is not too surprising, given that direct detection of the 
effects \cite{Katsnelson06b,KatsnelsonSSC,Cheianov06,Cheianov07} would require
an angle-resolved measurement and/or very clean samples.
Alternatively, one can ask if the behavior \cite{Katsnelson06b,KatsnelsonSSC,Cheianov06,Cheianov07}
can be inferred from the dependence of transport
properties on the magnetic field that often provides valuable insights
into electron dynamics.
It is interesting therefore to better understand
the signatures of ballistic transmission in external magnetic field,
which is the main purpose of the present work.

We start by noting that the coupling of an electron
to external 
fields reflects
relativistic character 
of charge carriers in graphene
with the speed of light $c$ replaced by $v_F$. 
In relativistic electro-magnetic theory the fields $\vec E$ and $\vec B$ 
are treated on equal footing, playing the role
of each other in a moving reference frame. The dynamics
of a relativistic particle in uniform fields 
depends only on the 
Lorentz invariants $\vec E^2-\vec B^2$, $\vec E.\vec B$ \cite{LL-2}.
In particular, the dynamics in crossed fields, $\vec E.\vec B=0$, 
can be of two main types, magnetic and electric,
depending on the relative strength of the fields $E$ and $B$. 
In the first case,
$B>E$, the particle trajectories are described by 
cyclotron motion superimposed with a drift perpendicular to $\vec E$. 
In the second case, 
$E>B$,
the trajectories are similar to those in the absence of $B$ field, 
moving asymptotically  parallel to 
$\vec E$ and exhibiting no cyclotron motion. 

Quantum transport in these two regimes, magnetic and electric, 
was discussed a while ago\,\cite{AronovPikus66,Weiler67} 
in the context of interband tunneling 
in two-band semiconductor systems 
modeled by the Dirac equation.
Naturally, both of these regimes can be realized
in graphene p-n junctions. In the magnetic case, realized 
for $B>(c/v_F)E$\,\cite{Lukose07}, electron motion is described
by quantized Landau levels with a linear dispersion
in the momentum perpendicular to $\vec E$, i.e. 
parallel to the junction. This defines relativistic Quantum Hall edge states\,\cite{MacDonald83}
transporting charge along the p-n interface. 
Cyclotron frequency in this regime as a function
of $E$ vanishes at $E=(v_F/c)B$,
signaling collapse of
the Landau levels and 
Quantum Hall effect\,\cite{Lukose07}.

In the electric regime $B<(c/v_F)E$, 
which will be of main interest for us here,
electrons can move freely along $\vec E$, transporting electric
current through the junction.
Transmission coefficient,
found below as 
a function of $B$,
is shown to vanish 
at the critical field
\be\label{eq:B*}
B=B_\ast \equiv (c/v_F)E.
\ee 
The effect of increasing magnetic field is therefore to pinch
off transport through the junction, and transform it into the edge 
state transport along the junction in the Quantum Hall state
at $B>B_\ast$. 
Similar conclusions for tunneling suppression by transverse magnetic field 
in 3D junctions modeled by Dirac particles with a finite mass were
obtained in Refs.\cite{AronovPikus66,Weiler67}.

\begin{figure}
\includegraphics[width=3.4in]{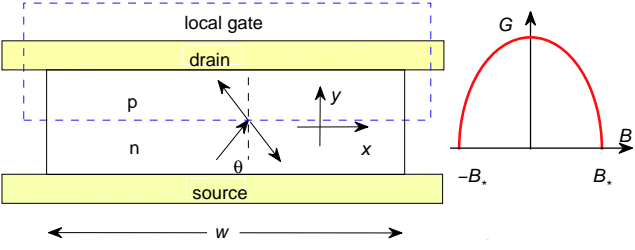}
\vspace{-0.85cm}
\caption[]{
Schematic of a p-n junction in a locally gated sample.
For the geometry shown (wide and short sample) the conductance
is dominated by the junction. Magnetic field suppresses conductance
as $G(B)\propto (1-(B/B_\ast)^2)^{3/4}$, Eq.(\ref{eq:G(B)}).
}
\label{fig1}
\end{figure}

In our approach, we solve the Dirac equation 
in crossed $E$ and $B$ fields exactly
with the help of a Lorentz boost. This allows us to treat the monolayer
and bilayer cases on equal footing. 
We find collimated 
transmission peaked at an angle $\sin\theta_B=B/B_\ast$, 
with unit transmission at the peak, $\theta=\theta_B$,
as in the absence of magnetic field. 
The net conductance, found by integrating
transmission over angles $\theta$, for a wide junction is given by
\be\label{eq:G(B)}
G(B\le B_\ast)=\frac{e^2}{2\pi h}\frac{w}{d}\lp 1-(B/B_\ast)^2\rp^{3/4}
,
\ee
where $d=(\hbar v_F/|eE|)^{1/2}$ and $w$ is the p-n interface length 
(see Fig.\ref{fig1}).
The suppression of tunneling (\ref{eq:G(B)}) precedes formation of 
edge states at the p-n interface at $B>B_\ast$.

To estimate the critical field $B_\ast$ for the parameter values 
of Refs.\cite{Huard07,Williams07,Ozyilmaz07} 
one would have to account for screening of the in-plane field
created by gates\,\cite{Fogler07}.
To bypass these complications,
we assume that a density variation of order 
$n_0\sim 10^{12}\,{\rm cm}^{-2}$
is created in a p-n junction across a distance $\ell\approx 50\,{\rm nm}$.
Then the field felt by the electrons is $eE\sim \hbar v_F\sqrt{\pi n_0}/\ell$,
giving 
\be\label{eq:estimate}
B_\ast =(c/v_F)E \sim (\hbar c/e)\sqrt{\pi n_0}/\ell
.
\ee
In terms of the magnetic length
$\ell_B=\sqrt{\hbar c/eB}$,
this  translates into 
$\ell_{B_\ast}^2=\ell/\sqrt{\pi n_0}\approx 260\,{\rm nm}^2$,
yielding an experimentally convenient value of $B_\ast\sim 2.5\,{\rm T}$.

For the p-n junction contribution to dominate over 
the conduction in the p and n regions, it is beneficial
to be in the ballistic regime, 
similar to Refs.\cite{Katsnelson06b,Cheianov06,Cheianov07},
and to use wide and short samples (see Fig.\ref{fig1}). 
These requirements are more relaxed for p-n junctions 
in epitaxial and bilayer systems, where tunneling
is exponentially suppressed owing to the presence of a spectral gap (see below).

We first consider transport in the p-n junction in the absence 
of magnetic field.
Massless Dirac particles in graphene
moving near the p-n interface in a uniform in-plane electric
field are described by the Hamiltonian
\be\label{eq:H_dirac}
H=e\phi(\vec x)+v_F\xi\left(
\begin{array}{cc}
0 & p_+ \\
p_- & 0 
\end{array}
\right)
,\quad
p_\pm=p_1\pm ip_2
,
\ee
where $\phi$ is the electrostatic potential used to create the junction,
and $\xi=\pm1$ for the points $K$ and $K'$.
We consider a p-n interface parallel to the $x$ axis (Fig.\ref{fig1}),
with the external field $\vec E\parallel \hat y$ 
described by $\phi(\vec x)=-Ex_2$.

The eigenstates of (\ref{eq:H_dirac})
are characterized by the momentum component parallel to the junction,
$\psi(t,\vec x)=e^{-i\epsilon t+ip_1x_1}\psi(x_2)$,
giving a 1D problem for $\psi(x_2)$.
Following Ref.\cite{KaneBlount}, we choose to write 
this problem
in momentum representation
\be\label{eq:p-representation}
-ieE\,d\psi/dp_2
= \tilde H\psi
,\quad
\tilde H= v_F(p_1\sigma_1-p_2\sigma_2) -\epsilon
.
\ee
As noted in Ref.\cite{KaneBlount}, momentum representation
provides direct access to the asymptotic scattering states,
and is thus more beneficial than the position representation.
 
Indeed, Eq.(\ref{eq:p-representation}), interpreted as a time-dependent 
evolution
with the Hamiltonian $\tilde H$, ``time'' $p_2$, and ``Planck's constant''
$eE$, 
can be identified with the Landau-Zener problem
for a two-level system evolving through an avoided crossing. 
Hence the
probability to be transmitted (reflected) 
in the Dirac problem translates into the probability of a diabatic
(adiabatic)
Landau-Zener transition.
The transmission coefficient can thus be found
using the answer for the latter\,\cite{LL-3}, giving
\be\label{eq:no_B}
T(p_1)=\exp(-\pi\hbar v_F p_1^2/|eE|)
,
\ee
which agrees with the results of \,\cite{KaneBlount,Cheianov06}
(see also \cite{Andreev07}).

Alternatively, the result (\ref{eq:no_B}) can be put in the context
of Klein tunneling that links transmission of a Dirac particle through
a steep barrier with electron/hole pair creation.
The pair creation rate can be found as the probability of 
an interband transition
occuring when the particle momentum evolves as $p_2 = eEt$.
Because each created pair
transfers one electron charge across the p-n interface,
the pair creation rate is equal to the tunneling current.

To analyze transport in the p-n junction in the presence of a magnetic field,
it will be convenient to rewrite
the Dirac equation (\ref{eq:H_dirac})
in a Lorentz-invariant form
\be\label{eq:dirac}
\gamma^\mu\lp p_\mu-a_\mu\rp \psi=0
, \quad
\{\gamma_\mu,\gamma_\nu\}_+=2g_{\mu\nu}
,
\ee
where $\gamma^\mu$ are Dirac gamma-matrices,
$\gamma^0=\sigma_3$, $\gamma^1=-i\sigma_2$, $\gamma^2=-i\sigma_1$,
and 
$\psi$ is a two-component wave function. 
Here we use the space-time notation for coordinates $x_\mu=(v_Ft,x_1,x_2)$,
momenta $p_\mu=\hbar(iv_F^{-1}\p_t,-i\p_{x_1},-i\p_{x_2})$,
and external field $a_\mu=(a_0,a_1,a_2)$.
The fields $\vec E\parallel \hat y$
and $\vec B\parallel \hat z$
are described by 
\be\label{eq:a_mu}
a_0=-\frac{e}{v_F}Ey,\quad
a_1=-\frac{e}{c}By,\quad 
a_2=0. 
\ee
The Dirac equation (\ref{eq:dirac}) is invariant under the Lorentz group 
($d=2+1$):
\bea
&& x^{\mu'}=\Lambda^{\mu'}_\mu x^\mu
, \quad
p_{\mu'}=\Lambda_{\mu'}^\mu p_\mu
, \quad
a_{\mu'}=\Lambda_{\mu'}^\mu a_\mu
, \quad
\\
&&
\psi'=S(\Lambda)\psi, 
\eea
where 
$S(\Lambda)=\exp\lp\frac18\omega_{\mu\nu}[\gamma^\mu,\gamma^\nu]\rp$ 
for $\Lambda=\exp(\omega)$.


We first find transmission quasiclasically, using 
the same factorization
as above,
$\psi(t,\vec x)=e^{-i\epsilon t+ip_1x_1}\psi(x_2)$,
which gives a 1D problem for $\psi(x_2)$:
\be\label{eq:1D_problem}
\lp \gamma^0(\epsilon+ax)+\gamma^1(p_1+bx)-i\gamma^2\p_x\rp \psi(x)=0
\ee
where $a=\frac{e}{v_F}E$, $b=\frac{e}{c}B$, $x\equiv x_2$.
Eq.(\ref{eq:1D_problem}) can be cast in the form of evolution
with a {\it non-hermitian}
Hamiltonian:
\be\label{eq:a,b}
i\p_x\psi(x)
=\lp (\epsilon+ax)\sigma_2+i(p_1+bx)\sigma_3\rp\psi(x)
.
\ee
Now, we apply the adiabatic approximation, 
constructed in terms of $x$-dependent 
eigenstates 
and eigenvalues of the non-hermitian
Hamiltonian. The eigenvalues are $\pm\kappa(x)$, where
$\kappa(x)=\sqrt{(\epsilon+ax)^2-(p_1+bx)^2}$.
This quantity is imaginary in the classically forbidden region $x_1<x<x_2$,
where $x_{1,2}=(\epsilon\pm p_1)/(a\pm b)$. The WKB transmission coefficient
then equals $e^{-S}$, where
\be\label{eq:Swkb}
S=2\int_{x_1}^{x_2}\Im \kappa(x)dx
=\pi\frac{(p_1 a-\epsilon b)^2}{(a^2-b^2)^{3/2}}
.
\ee
%
For $\vec B=0$ our WKB result (\ref{eq:Swkb}) agrees with Eq.(\ref{eq:no_B}).


The problem (\ref{eq:dirac}), (\ref{eq:a_mu}) can be solved 
exactly with the help of a Lorentz transformation
chosen so as to eliminate the field $\vec B$. 
(This is possible because the Lorentz-invariant combination $\vec B . \vec E$
equals zero.)
For a not too large magnetic field,
$B<B_\ast=\frac{c}{v_F}E$, we can eliminate $B$ by 
a Lorentz boost with velocity parallel to the junction:
\be\label{eq:L-boost}
\Lambda=\left(
\begin{array}{ccc}
\gamma & \gamma\beta & 0 \\
\gamma\beta & \gamma & 0 \\
0 & 0 & 1
\end{array}
\right)
,\quad \gamma=\frac1{\sqrt{1-\beta^2}}
\ee
Choosing the boost parameter as 
$\beta=-v_FB/cE=-B/B_\ast$, in the new frame we have $B'=0$,
$E'=E/\gamma$.

Because $B'=0$,
the transmission coefficient for 
an electron with momentum $p'_1$
parallel to the p-n junction 
is given by
$T=e^{ -\pi \hbar v_F {k'}_1^2/|eE'|}$ in the new frame (see Eq.(\ref{eq:no_B}) 
and \cite{KaneBlount,Cheianov06}). 
Expressing $p'_1$ and $E'$ through the quantities in the lab frame, we obtain
%
\be\label{eq:T(k,epsilon)}
T(p_1)= e^{ -\pi \gamma^3 d^2 (p_1+\beta\tilde\epsilon)^2}
,\quad
d= (\hbar v_F/|eE|)^{1/2}
,
\ee
$\tilde\epsilon=\epsilon/v_F$, which coincides with the WKB result (\ref{eq:Swkb}).

In passing from the moving and lab frames we used the fact
that the transmission coefficient $T$, 
Eq.(\ref{eq:T(k,epsilon)}), 
is a scalar with respect to Lorentz transformations (\ref{eq:L-boost}).
This is true because transmission and reflection
at the p-n interface
is interpreted in the same way
by all observers moving 
with the velocity parallel to the interface. 

\begin{figure}
\includegraphics[width=3.4in]{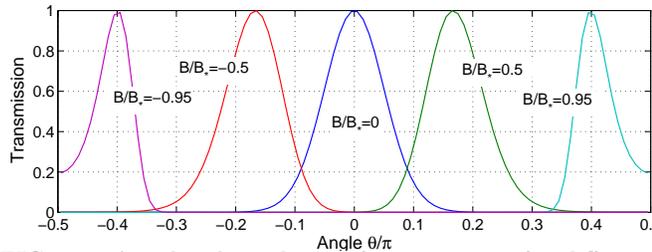}
\vspace{-0.85cm}
\caption[]{
Angular dependence of transmission for different magnetic field values, $T(\theta)=e^{-\alpha\gamma^3(\sin\theta-B/B_\ast)^2}$, Eq.(\ref{eq:T(k,epsilon)}), for $\alpha=\pi (d/\lambdabar_F)^2=20$.
Transmission reaches unity at a field-dependent angle 
$\theta_B=\arcsin B/B_\ast$.
}
\label{fig2}
\end{figure}

The dependence of the transparency (\ref{eq:T(k,epsilon)})
on the electric field $E$ 
is such that $T$ grows as $E$ increases. This is a manifestation 
of the Klein tunneling phenomenon 
in which steeper barriers yield higher transmission.

\begin{figure}
\includegraphics[width=3.6in]{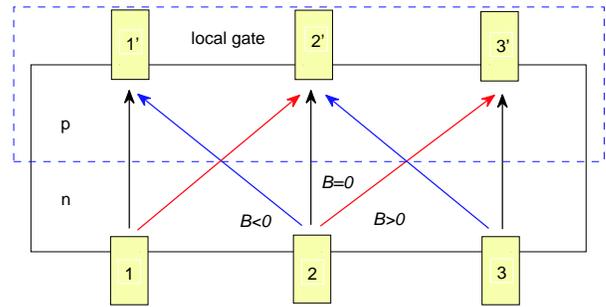}
\vspace{-0.85cm}
\caption[]{
Field-controled switching
of collimated current flow through a ballistic p-n junction between different 
contacts.
}
\label{fig3}
\end{figure}

The result (\ref{eq:T(k,epsilon)}) features exponential suppression 
of tunneling by $B$ field
for all momenta except $p_1=-\beta\epsilon$ that yields perfect transmission.
This corresponds to the incidence angle $\theta_B=\arcsin B/B_\ast$
(see Fig.\ref{fig2}).
At equal p and n densities, the velocities of transmitted particles are collimated at 
$\theta\approx \theta_B$, with the collimation angle variance
determined by
$\Delta p_1\approx d^{-1}\gamma^{-3/2}$.
This gives an estimate
\be\label{eq:theta_B}
\Delta\theta\sim (\lambdabar/d)\lp 1-(B/B_\ast)^2\rp^{1/4}
,\quad
\lambdabar=v_F/\epsilon_F
.
\ee
We conclude that the nearly unit transmission,
which occurs perpendicular to the p-n interface at $B=0$\,\cite{Katsnelson06b,Cheianov06}, persists
at finite magnetic fields, albeit for $\theta_B\ne 0$.
This behavior of the collimation angle can be used to
realize a switch (see Fig.\ref{fig3}),
in which current is channeled between diferent pairs of contacts
by varying the $B$ field.

The p-n junction net conductance can be found from the Landauer formula
\be\label{eq:Landauer_formula}
G=\frac{e^2}{h}\sum_{-k_F<p_1<k_F}T(p_1)=\frac{w e^2}{2\pi h}
\int_{-k_F}^{k_F}T(p_1)dp_1
\ee
where $w$ is the length of the junction interface (see Fig.\ref{fig1}),
and the states contributing to transport are those at the Fermi level,
$\epsilon=\epsilon_F$. 
For a wide junction, $w\gg d\gg \lambdabar_F$, 
extending integration over $p_1$ to infinity we obtain
the $\lp 1-(B/B_\ast)^2\rp^{3/4}$ dependence (\ref{eq:G(B)}).

It is interesting to apply these results to epitaxial
graphene, described by massive Dirac particles 
$\epsilon=\pm(v_F^2\vec p^2+\Delta^2)^{1/2}$
with an energy gap $\Delta$
induced by the substrate\,\cite{Lanzara,Mattausch07}. 
The generalization amounts to
replacing $p_1^2$ by $p_1^2+\Delta^2/v_F^2$
in (\ref{eq:no_B}).
Performing Lorentz transformation,
we find exponential suppression of conductance:
\be\label{eq:G(B)_gapped}
G(B)=\frac{e^2}{2\pi h}\frac{w}{d}\lp 1-\beta^2\rp^{3/4}
\textstyle{
\exp\lp -\frac{\pi d^2 \Delta^2}{v_F^2(1-\beta^2)^{1/2}} \rp
}
\ee
({\it cf.} Refs.\cite{AronovPikus66,Weiler67}).
The angular dependence of transmission in this case is the same as in the
massless case.

We note that $G(B\ge B_\ast)=0$ does not necessarily mean that the system ceases
to conduct. The behavior predicted by Eq.(\ref{eq:G(B)})
at $B\ge B_\ast$ should be interpreted as 2D transport pinching off  
by the onset of the Quantum Hall effect. 
In that, just the part of the conductance proportional to the sample width $w$
vanishes, while the edge mode contribution remains nonzero.

\begin{figure}
\includegraphics[height=2in]{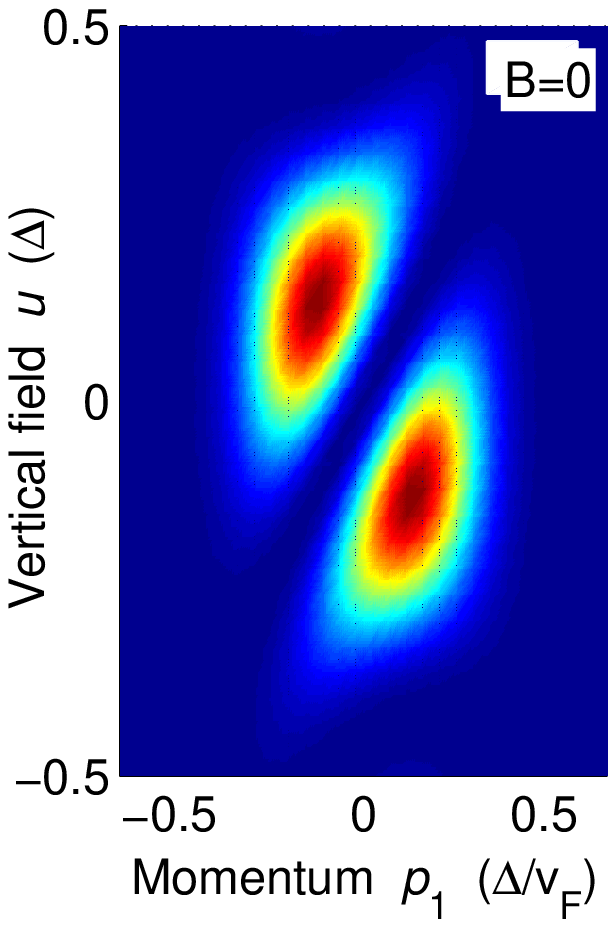}
\includegraphics[height=2in]{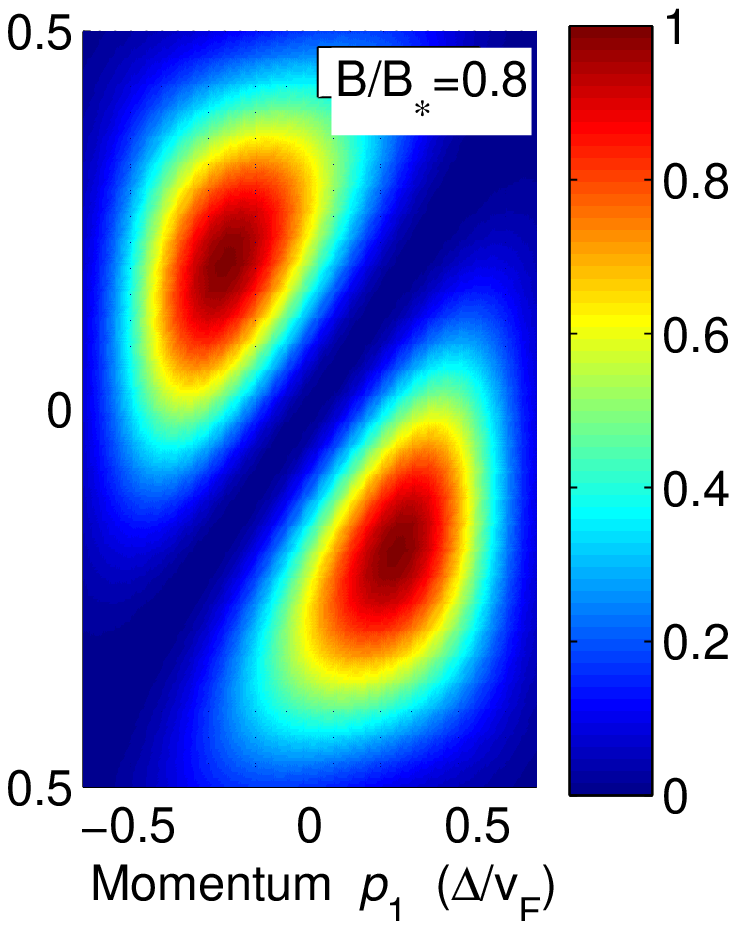}
\vspace{-0.5cm}
\caption[]{
Transmission of a bilayer p-n junction
{\it vs.} momentum 
parallel to the interface and vertical field.
Note the double hump structure with
unit transmission at the peak,
and suppression of 
tunneling at large $u$ and $p_1$ 
($E'=0.05\Delta^2/ev_F$).
}
\label{fig4}
\end{figure}

Our approach can be readily generalized to described 
recently fabricated p-n junctions in 
graphene bilayers\,\cite{Oostinga07}. 
The bilayer Hamiltonian\,\cite{McCann06} includes the standard monolayer 
tight-binding
part, as well as 
a direct coupling between the adjacent sites $B$, $\tilde A$
of different monolayers
and a weaker coupling between non-adjacent sites $A$, $\tilde B$:
$\gamma_{\tilde B A}\ll\gamma_{B\tilde A}\approx 0.6\,{\rm eV}$ in notation 
of Ref.\cite{McCann06}. Here, for simplicity, 
we ignore $\gamma_{\tilde B A}$ and denote $\gamma_{B\tilde A}$
as $\Delta$.

It is convenient to write 
the bilayer Hamiltonian, linearized near the Dirac points,
in pseudospin notation, using 
$\tau_3=\pm 1$ to
label the monolayers. The inter-layer coupling 
takes the form
$H_\Delta=\Delta\lp \tau_+\sigma_-+\tau_-\sigma_+\rp
=\frac{\Delta}4\lp \tau_1\sigma_1+\tau_2\sigma_2\rp$, 
where $\sigma_\pm=\frac12(\sigma_1\pm i\sigma_2)$,
$\tau_\pm=\frac12(\tau_1\pm i\tau_2)$. 
This gives the Hamiltonian
%
\be
H = v_Fp_1\sigma_1-v_Fp_2\sigma_2+\frac12 u\tau_3+
\frac{\Delta}2\lp \tau_1\sigma_1+\tau_2\sigma_2\rp
,
\ee
where $u$ is the vertical field that opens a gap of size $|u|$
in the bilayer spectrum. 
Multiplying the time-dependent Schr\"odinger equation by $\sigma_3$,
we rewrite it
as a Dirac equation (\ref{eq:dirac})
with a fictitious $\tau$-dependent gauge field:
\be
\gamma^\mu (p_\mu-a_\mu-g_\mu)\psi=0
,\quad
g_\mu=\lp \tilde u\tau_3, -\tilde \Delta\tau_1,\tilde \Delta\tau_2\rp
,
\ee
where $\tilde u=u/2v_F$, $\tilde \Delta=\Delta/2v_F$, and
the external field $a_\mu$ is defined in the same way as above.

Under Lorentz boost (\ref{eq:L-boost})
the equation $\gamma^\mu (p_\mu-a_\mu-g_\mu)=0$ 
changes covariantly with the momenta and fields transforming 
via $p'=\Lambda p$, $a'=\Lambda a$, $g'=\Lambda g$,
giving $g_{\mu'}=\frac1{2v_F}\lp \gamma(u\tau_3-\beta\Delta\tau_1),
\gamma(\beta u\tau_3-\Delta\tau_1),\Delta\tau_2\rp$.
%
Choosing $\beta$ so as to eliminate the $B$ field, 
we find the transformed Hamiltonian
$H'=-eE'x_2+H_k(p'_1,p'_2)$, where 
%
\bea\label{eq:H'}
&& H_k(p'_1,p'_2)=\frac1{2}\gamma\lp u\tau_3-\beta\Delta\tau_1\rp 
+
\\\nonumber
&&
\lp v_Fp'_1-\frac12\gamma(\beta u\tau_3-\Delta\tau_1)\rp\sigma_1
-\lp v_Fp'_2-\frac12\Delta\tau_2\rp\sigma_2
.
\eea
Working in the momentum representation, as above, we treat $\epsilon'\psi'=H'\psi'$
as a first-order differential equation
\[
ieE'\,d\psi/dp'_2=\lp H_k(p'_1,p'_2)-\epsilon'\rp\psi
.
\]
%
We evaluate the transfer matrix of this equation numerically, 
and find that in the physically interesting case $u\ll\Delta$, 
the lowest and the uppermost energy levels of $H_k$ are decoupled 
from the two middle levels. The $4\times4$ transfer matrix 
is thus reduced to a $2\times2$ matrix, 
yielding the transmission and 
reflection coefficients.

Transmission features an interesting behavior
as a function of external fields and particle momentum (see Fig.\ref{fig4}).
It has a symmetric 
double hump profile as a function of $p_1$ and $u$
vanishing between the humps (unlike single gaussian peak
in the monolayer case) and, somewhat unexpectedly,
perfect transmission at the peak.
At large $p_1$ and $u$, because of the energy gap
opening, transmision is strongly suppressed.  
Conductance, found from the Landauer formula (\ref{eq:Landauer_formula}),
also exhibits strong suppression at increasing $u$
and $B$, qualitatively similar to the gapped 
monolayer case, Eq.(\ref{eq:G(B)_gapped}).


We benefited from useful discussions
with C. M. Marcus and D. A. Abanin.
This work is supported by the DOE
(contract DEAC 02-98 CH 10886),
NSF MRSEC (DMR 02132802) and
NSF-NIRT DMR-0304019.

\end{document}